\documentclass[%
 reprint,
superscriptaddress,
 amsmath,amssymb,
 aps,
pre,
]{revtex4-2}

\usepackage{epsfig,dsfont,amssymb,amsmath,amsthm,amsfonts,amsbsy,mathrsfs}
\usepackage{graphicx}
\usepackage{color}
\usepackage{bm}
\usepackage{multirow}
\usepackage{natbib}

\newcommand{\BEQ}{\begin{equation}}
\newcommand{\EEQ}{\end{equation}}
\newcommand{\BEA}{\begin{eqnarray}}
\newcommand{\EEA}{\end{eqnarray}}

\usepackage{tikz}
\usetikzlibrary{arrows,shapes,chains}

\newcommand{\rev}[1]{\textcolor{black}{#1}}

\begin{document}
\preprint{APS/123-QED}

\title{Scaling of the entropy production rate in a $\varphi^4$ model of Active Matter}

\author{Matteo Paoluzzi}
\email{matteopaoluzzi@ub.edu}
\affiliation{Departament de Física de la Mat\`eria Condensada, Universitat de Barcelona, C. Martí Franqu\`es 1, 08028 Barcelona, Spain.}

\date{\today}

\begin{abstract}
In active $\varphi^4$ field theories the nonequilibrium terms play an important role in describing active phase
separation; however, they are irrelevant, in the renormalization group sense, at the critical point. Their irrelevance
makes the critical exponents the same as those of the Ising universality class. Despite their irrelevance, they
contribute to a nontrivial scaling of the entropy production rate at criticality. We consider the nonequilibrium
dynamics of a nonconserved scalar field $\varphi$ (Model A) driven out-of-equilibrium by a persistent noise that is
correlated on a finite timescale $\tau$, as in the case of active baths. We perform the computation of the density of
entropy production rate $\sigma$ and we study its scaling near the critical point. We find that similar to the case of
active Model A, and although the nonlinearities responsible for nonvanishing entropy production rates in the two
models are quite different, the irrelevant parameter $\tau$ makes the critical dynamics irreversible.
\end{abstract}

\maketitle

\section*{Introduction} \label{sec:Intro}
Self-propelled particles can undergo a second-order phase transition whose numerical estimate of the critical exponents are consistent with those of the Ising universality class \cite{PhysRevLett.123.068002,maggi2021universality}. 
This happens despite the microscopic mechanism responsible for this non-equilibrium phase transition, i.e., the so-called Motility-Induced Phase Separation, does not have any equilibrium counterpart \cite{Tailleur08,cates2015motility}. 
Coarse-grained models of active particles suggest that non-equilibrium gradient terms have to be taken into account for describing the active system on a large scale, e.g., the so-called Active Model A, Active Model B, etc... \cite{cates2019active}. However, once we study the system around the critical point, 
the non-equilibrium terms of active $\varphi^4$ field theories are irrelevant in the Renormalization Group (RG) sense and thus they run to zero under RG transformations \cite{PhysRevLett.124.240604}. It has been shown recently that, even though irrelevant, the non-equilibrium gradient term $(\nabla \varphi)^2$ of Active Model A/B causes an anomalous scaling of the density of entropy production rate $\sigma$. This result suggests that non-equilibrium irrelevant terms have to be carefully studied around the critical point.
 We can speculate that this fact might open the way to the definition of new non-equilibrium universality classes.

On the other hand, one can expect that the non-equilibrium terms considered in Active Model A/B are not the only ingredients that one should take into account for 
building an active $\varphi^4$ field theory. And thus, a natural question is whether or not other non-equilibrium terms make the critical dynamics effectively non-equilibrium.
Usually, one assumes that the noise acting on the coarse-grained fields has no memory, not in time, nor in space. This assumption seems reasonable because, to describe the system on a large scale, one has to maintain only the relevant degrees of freedom, e.g., the order parameters, which are slow-varying variables. In this framework, the noise represents the effect of the fast degrees of freedom on the slow ones \cite{hohenberg1977theory}.

In the case of active systems, the noise is not necessarily delta-correlated, as proved by experiments of passive beads in active baths \cite{PhysRevLett.84.3017,Maggi14,maggi2017memory}.
Moreover, not only is the noise generated by the active bath correlated on a finite time-scale $\tau$, but also it violates the Fluctuation-Dissipation Theorem in a peculiar way \cite{maggi2017memory}, i.e., although the active bath exerts time-correlated forces, linear response reveals an instantaneous friction kernel.

If we interpret the passive bead as a probe that captures the emerging properties of the active bath, it is reasonable to assume that fluctuations
are not delta-correlated but they are characterized by a finite correlation time. 
Moreover, recent numerical experiments suggest that even approaching the MIPS critical point, i.e., in a situation where the critical slowing down dominates over any other time scale, field theories driven by an exponentially-correlated stochastic force well capture the non-thermal {\it active} fluctuations \cite{Maggi2021}.
It is known that, once we consider an exponentially correlated noise in a system composed of discrete degrees of freedom, a non-vanishing entropy production rate can be due only to non-linear interactions \cite{caprini2019entropy,Fodor16}. A natural question is what happens in the case of a field theory.

In this work, using the standard machinery of dynamical critical phenomena and stochastic thermodynamics, we compute the entropy production rate in the case of Model A driven by a persistent noise on a finite time scale $\tau$. We focus our attention on the behavior of the system at the critical point and we consider the situation where the transition is approached from the symmetric phase, i.e., the system is not phase-separated, and thus there are no reasons for including other non-equilibrium {\it active} terms. 
The computation does not require any perturbation expansion neither in the persistence 
time $\tau$ nor in the non-linear couplings of the field theory.

We obtain that all the non-linear terms (in principle infinite) appearing in the effective Hamiltonian that describes the critical system contribute to the entropy production. Moreover, focusing our attention on a $\varphi^4$ theory, we show that, in agreement with that observed in Active Model A \cite{PhysRevLett.124.240604}, the irrelevant parameter $\tau$ provides a source of entropy production at the critical point.

\section*{Model A driven by a persistent noise}
As it has been recently shown in Ref. \cite{maggi2021universality} (where one can find details about the RG of Model B driven by a persistent noise), field theories with exponentially correlated noise well reproduce the critical dynamics of MIPS and are consistent with the Ising universality class.
Here we perform the computation of the entropy production rate to understand the impact of the persistent noise on the critical dynamics. Following Ref. \cite{hohenberg1977theory}, we focus our attention on the simplest case of Model A. 
We consider the non-equilibrium dynamics of a non-conserved scalar field $\varphi=\varphi(x,t)$ in $d$ spatial dimensions (with $x\equiv \mathbf{x} \in \mathbb{R}^d$) that is driven out-of-equilibrium by a persistent stochastic force $\eta=\eta(x,t)$ that is correlated on a finite time scale $\tau$. The stochastic dynamics of the two fields are  
\begin{align} \label{eq:dyn_ModelA_1}
    \dot{\varphi} &= -\Gamma_0\frac{\delta H_{LG}}{\delta \varphi}  + \eta \\ \label{eq:dyn_ModelA_2}
\dot{\eta}    &= -\frac{1}{\tau} \eta + \zeta \; ,
\end{align}
with the noise $\zeta = \zeta(x,t)$ satisfying 
\begin{equation} \nonumber
    \langle \zeta(x,t) \rangle = 0 \; , \;\; \langle \zeta(x,t) \zeta(y,s)\rangle = \frac{2 D \Gamma_0}{\tau^2} \delta(t-s) \delta^{(d)}(x-y) \; .
\end{equation}
In the present work 
we set $\Gamma_0\!=\!1$.
We now consider the most general expression for the Landau-Ginzburg Hamiltonian $H_{LG}$ which is as follows
\begin{align} \label{eq:HLG}
    H_{LG}[\varphi] &= \int dx\, \left( H_0 + H_I \right) \\ \nonumber 
    H_0 &\equiv  \frac{1}{2} \left[ \mu (\nabla \varphi)^2 + r \varphi^2 \right] \\ \nonumber 
    H_I &\equiv  \sum_{p=2}^{\infty} \frac{u_p}{(2p)!}\varphi^{2p} \; ,
\end{align}
where we adopt the notation $dx \equiv d^dx$. In Eq. (\ref{eq:HLG}) $H_0$ is the Gaussian contribution and $H_I$ contains the non-linear interactions. The coupling constant $r$ sets the distance from the \rev{mean-field} critical point that is located at $r\!=\!0$. 
 We work under a condition such that the critical point is always approached from 
 from the disordered phase where $r \geq 0$ and thus $\langle \varphi \rangle = 0$. 
We stress that, since we are interested in studying only the contribution of the persistent noise on the entropy production rate around the critical point, we neglect other non-equilibrium terms like those in Active Model A and B \cite{cates2019active}. These terms surely
play an important role in active phase separation and thus 
have to be taken into account for describing the spinodal decomposed active phases \cite{wittkowski2014scalar,stenhammar2013continuum}. 
To establish the relevant parameters of the theory, we can perform the scaling analysis of Eq. (\ref{eq:dyn_ModelA_1}) that is obtained by considering Kadanoff transformations with the scaling parameter $b$; i.e., we replace $x \to b x$, $t \to b^z t$, and $\varphi \to b^\chi \varphi$ (more details about the scaling analysis of the dynamical action can be found in Ref. \cite{Maggi2021}).
We obtain  the following scaling transformations for the parameters of the theory 
\begin{eqnarray} \label{eq:scaling}
r^\prime    &=& r b^z\\ \nonumber
\mu^\prime  &=& \mu b^{z-2}\\ \nonumber
u^\prime_p  &=& u_p b^{2(p-1) \chi + z}\\ \nonumber
D^\prime    &=& D b^{z - 2 \chi - d}\\ \nonumber
\tau^\prime &=& \tau b^{-z} \; ,
\end{eqnarray}
where $z$ is the dynamic exponent and $\chi$ is the dimension of the scalar field $\varphi$.
Once we impose $\mu = \mu^\prime$, we get $z\!=\!2$, i.e., the usual $z$ exponent of Model A \cite{hohenberg1977theory}. 
Because of that, $\tau$ is an irrelevant parameter that runs to zero under RG transformations \cite{sancho1998non,Maggi2021} and thus the critical exponents are the same as those of the Ising universality class. However, non-universal quantities as the location of the critical point might depend on $\tau$, as shown in Ref. \cite{Paoluzzi16}.
From the scaling of the noise strength $D$ we obtain that the natural dimension of the field is $\chi=1 - d/2$, i.e., which is obtained by setting $D\!=\!D^\prime$. The scaling dimension of the couplings constants $u_p$ sets the (upper) critical dimensions of each non-linear interaction, i.e., $d_c=4$ for $u_2$ (which we name $u$), $d_c = 3$ for $u_3$, etc... meaning that the $\varphi^4$ interaction can be neglected above $4$ and $\varphi^6$ interactions can be neglected above $d=3$ \cite{ma2018modern}. 
As discussed in Ref. \cite{PhysRevLett.124.240604}, non-equilibrium control parameters might play a role at the critical point even if they are irrelevant in the RG sense. To study the impact of $\tau$ on critical dynamics, we compute the entropy production rate of the model.

Following stochastic thermodynamics \cite{seifert2012stochastic,lebowitz1999gallavotti,nardini2017entropy}, we compute the steady-state entropy production rate $\mathcal{S}$
\begin{align} \label{eq:epr}
\mathcal{S} &= \lim_{t_F \to \infty } \frac{1}{t_F} \left\langle \log \frac{P[\varphi]}{P_R[\varphi]} \right\rangle \; .
\end{align}
with $P[\varphi]$ indicating the probability of a path $\{ \varphi(x,t) \}_{t_0 \leq t \leq t_F}$, with $t_0$ and $t_F$ being the initial time and final time, respectively. 
In Eq. (\ref{eq:epr}), $P_R[\varphi]$ is the probability for the time-reversed trajectory that can be computed considering the transformations $t \to t_F - t$ and $\varphi \to \varphi_R(x,t) = \varphi(x,t_F - t) $.
The average $\langle \cdot \rangle$ is performed with respect to the noise realizations. Under suitable ergodicity assumptions that we make in the present work, when it is necessary, we replace averages over independent noise realizations with the average over a long trajectory.  

As we shall see, the probability $P[\varphi]$ can be written in terms of a dynamical action $A[\varphi]$, i.e., the Onsager–Machlup action, so that
\begin{eqnarray} \label{eq:OM}
P[\varphi] &\propto& \exp{ \left(- A[\varphi] \right) } \\
A[\varphi] &=& \int dx\,dt\; \mathcal{L}[\varphi] \; .
\end{eqnarray}
Here we have written the dynamical action $A[\varphi]$ in terms of the "density" $\mathcal{L}[\varphi]$.
Similarly for $P_R[\varphi]$ one has $P_R[\varphi] \propto \exp{ \left(- A_R[\varphi] \right) }$ so that
\begin{align}
    \mathcal{S} &= \lim_{t_F \to \infty} \frac{1}{t_F}\int_{t_0}^{t_0+t_F} dt\,dx\, \left\{ \mathcal{L}_R[\varphi] - \mathcal{L}[\varphi]\right\} \; .
\end{align}

\section*{Onsager-Machlup dynamical action}
To proceed with the program illustrated above,
it is convenient to write the dynamics in the Fourier space. We consider the system closed in a $d-$dimensional box of side $L$ and the volume is $V=L^d$ where periodic boundary conditions are employed. We can thus write $\varphi(x,t)$ and $\eta(x,t)$ in terms of their space-Fourier components as follows \cite{amit2005field,ma1973introduction,ma2018modern}
\begin{eqnarray}
    \varphi(x,t) &=& \frac{1}{L^{d/2}}  \sum_{ |k| \in (0,\Lambda) } e^{i k x} \varphi_k(t) \\ \nonumber
        \eta(x,t) &=& \frac{1}{L^{d/2}}  \sum_{ |k| \in (0,\Lambda) } e^{i k x} \eta_k(t)
\end{eqnarray}
where we have introduced the ultraviolet cutoff $\Lambda = \frac{2 \pi }{a}$ and $a$ is the lattice spacing (or the typical particle size) used as a regulator.
In the Fourier space, the Hamiltonian is
\begin{align}
    H_{LG} &= H_0 + H_I \\ \nonumber 
    H_0 &= \frac{1}{2} \sum_k \left( \mu k^2 + r\right) |\varphi_k|^2 \\ \nonumber 
    H_I &= \sum_{p=2}^{\infty}\frac{u_p}{(2p)!} \frac{1}{V^{-p+1} }\sum_{k_1,...,k_{2p}} \varphi_{k_1} \varphi_{k_2} \dots \varphi_{k_{2p}} \delta_{k_1  + \dots + k_{2p}} \; 
\end{align}
where $H_I$ contains all the non-linear interactions.
In this way, the dynamics is as follows:
\begin{eqnarray} \label{eq:phik}
\dot{\varphi}_k &=& -\frac{\partial H_{LG}}{ \partial \varphi_k} + \eta_k \\ \label{eq:etak}
\dot{\eta}_k    &=& -\frac{1}{\tau} \eta_k + \zeta_k
\end{eqnarray}
with 
\begin{align*}
    \langle \zeta_k(t) \rangle = 0 
    ,  \; \langle \zeta_k(t) \zeta_q(s) \rangle = \frac{2 D}{\tau^2} \delta_{k,-q} \delta(t-s) \; .
\end{align*}
For making progress, we perform the time derivative of Eq. (\ref{eq:phik}) obtaining
\begin{align} \label{eq:referee_coglione}
    \ddot{\varphi}_k &+ \mathbb{H}_{kq} \dot{\varphi}_q + \frac{1}{\tau} \left[ \dot{\varphi}_k + F_k\right] = \zeta_k \;  \\ \nonumber 
    F_k &\equiv \frac{\partial H_{LG}}{\partial \varphi_k} \\ \nonumber 
    \mathbb{H}_{kq} &\equiv \frac{\partial^2 H_{LG}}{\partial \varphi_k \partial \varphi_q} \; ,
\end{align}
where we have adopted the Einstein summation convention.
In order to obtain the expression of the probability distribution $P[\varphi]$, we start with introducing the expectation value $\langle \mathcal{O} \rangle$ of a generic observable $\mathcal{O}[\varphi]$, which is
\begin{align}
\langle \mathcal{O} \rangle = \int \mathcal{D}[\zeta_k] \mathcal{D}[\varphi_k] \, \mathcal{P}[\zeta] \mathcal{O}[\varphi] \delta \left[ \varphi - \varphi_\zeta \right]    
\end{align}
with $\varphi_\zeta$ indicating a solution of the Langevin equation (\ref{eq:referee_coglione}) and $\mathcal{P}[\zeta]$ the Gaussian distribution of the noise $\zeta_k$. We now perform some standard manipulations \cite{tauber2014critical,PhysRevA.8.423,PhysRevB.18.353,jensen1981functional}. First, we perform the change of variable 
\begin{align*}
    \delta \left[ \varphi - \varphi_\xi \right] &= \mathcal{J}[\varphi] \delta\left[ \ddot{\varphi}_k + \mathbb{H}_{kq} \dot{\varphi}_q + \frac{1}{\tau} \left[ \dot{\varphi}_k + F_k\right] - \zeta_k \right] \\
    \mathcal{J}[\varphi] &\equiv |\det \frac{\delta \xi_k(t)}{\delta \varphi_q(t^\prime)}| \; .
\end{align*}
Second, we represent the delta-functional through a set of response fields $\hat{\varphi}_k$. Third, we perform the Gaussian integral over the noise $\zeta_k$. In the following, we neglect the contribution of $\mathcal{J}[\varphi]$ since it does not contribute to the entropy production rate \cite{jac1,jac2}.
These standard manipulations \cite{PhysRevA.8.423,PhysRevB.18.353,jensen1981functional} lead to
\begin{align}
\langle \mathcal{O} \rangle &= \int \mathcal{D}[\varphi_k] \mathcal{D}[\hat{\varphi}_k] \, e^{-S[\hat{\varphi},\varphi]} \mathcal{O}[\varphi] \; ,
\end{align}
where we $S[\hat{\varphi},\varphi]$ is the so-called Janssen-De Dominicis response functional \cite{tauber2014critical,dominicis1976technics,janssen1976lagrangean}
\begin{eqnarray}  \label{eq:JD}
S[\hat{\varphi},\varphi] &\equiv& \frac{D}{2 \tau^2} \sum_k \int_{t_0}^{t_F} dt \hat{\varphi}_k^2 \\ \nonumber
&-& \frac{1}{\tau} \sum_k \int_{t_0}^{t_F} dt \hat{\varphi}_k \left[ \tau \ddot{\varphi}_k + M_{kq} \dot{\varphi}_q + F_k \right]  \\ \nonumber 
M_{kq} &\equiv& \delta_{kq} + \tau \frac{\partial^2 H_{LG}}{\partial \varphi_k \partial \varphi_q}\; .
\end{eqnarray}
After performing the Gaussian integration over $\hat{\varphi}_k$, we finally obtain that the probability $P[\varphi]$ of a path $\{ \varphi_k(t) \}_{t_0 \leq t \leq t_F}$ can be expressed in terms of the following Onsager-Machlup action (the proportional symbol indicates that we are neglecting factors that do not contribute to the computation of the entropy production rate)
\begin{eqnarray}
P[\varphi] &\propto& \exp{\left(-A[\varphi]\right)} \\ \nonumber 
A[\varphi] &=& \frac{1}{2D} \sum_k \int_{t_0}^{t_F} dt \, \left[ \tau \ddot{\varphi}_k + M_{kq} \dot{\varphi}_q + F_k \right]^2 \; .
\end{eqnarray}
Since we are interested in the critical behavior of the system by approaching the critical point 
from the {\it disordered phase} where $\langle \varphi \rangle=0$, the dynamical action $A[\varphi]$ is well-defined for every value of $\tau$. In such a situation the Hessian matrix $\mathbb{H}_{kq}$ is positive-definite and so $M_{kq}$ is positive-definite as well.

The probability of the time-reversed path can be obtained by employing the time-reversal operator $t \to t_F - t$ that brings us to the following expressions
\begin{eqnarray}
P_R[\varphi] &\propto& \exp{\left(-A_R[\varphi]\right)} \\  \nonumber
A_R[\varphi] &=& \frac{1}{2D} \sum_k \int_{t_0}^{t_F - t_0} dt \, \left[ \tau \ddot{\varphi}_k - M_{kq} \dot{\varphi}_q + F_k \right]^2 \, .
\end{eqnarray}

\section*{Entropy Production Rate}
Now we can perform the computation of the entropy production rate using the expression
\begin{eqnarray} \nonumber
    S &=& \lim_{t_F \to \infty} \frac{1}{2 D t_F} \sum_k \int_{t_0}^{t_F - t_0} dt \left\{   
    \left[ \tau \ddot{\varphi}_k + M_{kq} \dot{\varphi}_q + F_k \right]^2
    \right. \\ 
&-&    \left[ \tau \ddot{\varphi}_k - M_{kq} \dot{\varphi}_q + F_k \right]^2
    \left.
        \right\} \; ,
\end{eqnarray}
which brings us to
\begin{align}
    \mathcal{S} &= \lim_{t_F \to \infty } \frac{1}{D t_F} \sum_q \int_{t_0}^{t_F-t_0} dt \, \left[ A_q + B_q \right] \\ \nonumber 
    A_q &\equiv \dot{\varphi}_q \left[ \tau \ddot{\varphi}_q + F_q +  \tau \mathbb{H}_{kq} F_k \right] \\ \nonumber 
    B_q &\equiv \tau^2 \ddot{\varphi}_k \mathbb{H}_{kq} \dot{\varphi}_q \; .
\end{align}
The term $A_q$ can be written as
\begin{align}
   A_q &= \frac{d}{dt} Q_q \\ \nonumber 
   Q_q &= \frac{\tau}{2} \dot{\varphi}_q^2 + H_{LG} + \tau F_q \; .
\end{align}
The time integration of $A_q$ is a border term and thus vanishes in the limit $t_F\to\infty$ because of the factor $t_F^{-1}$. After integrating by parts $B_q$, we finally get 
\begin{eqnarray}
\mathcal{S} &=& -\lim_{t_F \to \infty} \frac{\tau^2}{D t_F}\int_{t_0}^{t_F - t_0} dt \, \mathbb{H}_{kq} \ddot{\varphi}_k \dot{\varphi}_q \; .
\end{eqnarray}
Performing an integration by parts and neglecting vanishing boundary terms, we obtain the following expression for the entropy production rate
\begin{eqnarray} \label{eq:sigma1}
\mathcal{S} &=& \frac{\tau^2}{2 D} \left \langle \dot{\varphi}_k \dot{\varphi}_q \dot{\varphi}_l G_{kql}\right \rangle \\ \label{eq:sigma2}
G_{kql} &\equiv& \frac{\partial^3 H_{LG}}{\partial \varphi_k \partial \varphi_q \partial \varphi_l } = \frac{\partial^3 H_I}{\partial \varphi_k \partial \varphi_q \partial \varphi_l}\; .
  \end{eqnarray}
This is a central result of this paper: From Eqs. (\ref{eq:sigma1}) and (\ref{eq:sigma2}) we immediately realize that, in principle, all the non-linear terms in a Landau-Ginzburg model contribute to the entropy production. 
Moreover, the expression obtained here generalized to field theories the entropy production rate formula obtained in the case of Active Ornstein-Uhlenbeck particles \cite{Fodor16,caprini2019entropy}. Finally, no matter how big $\tau$ is, we notice that for a Gaussian model one has $\mathcal{S}=0$. Using Eqs. (\ref{eq:sigma1}) and (\ref{eq:sigma2}), $\mathcal{S}$ can be computed numerically by solving the equations of motion. This can be done, for instance, by discretizing the dynamics on a $d-$dimensional grid.

Once we perform the inverse Fourier transform, we get the following expression
\begin{align} \label{eq:epr_new}
    \mathcal{S} &= \int dx \, \sigma(x) \\ \nonumber 
    \sigma &\equiv \frac{\tau^2}{2 D}\left\langle \dot{\varphi}(x)^3 \frac{\delta^3 H_I}{\delta \varphi(x)^3} \right\rangle \; ,
\end{align}
where $\sigma$ is the density of the entropy production rate.

\subsection*{The critical scaling}
We now study the scaling of $\sigma$ close to the critical point in the case of the $\varphi^4$ theory. 
For doing that, since $\varphi$ satisfies the equation of motion (\ref{eq:dyn_ModelA_1}), we can write 
\begin{align}
    \sigma=\frac{\tau^2}{2 D} \left\langle (-\frac{\delta H_{LG}}{\delta \varphi} + \eta)^3 \frac{\delta^3 H_{LG}}{\delta \varphi^3} \right\rangle_\eta \; .
\end{align}
Different from Active Model A, where averages involve stochastic trajectories of $\varphi$ driven by a delta-correlated noise, here the stochastic force $\eta$ is time-correlated. Because of that, we do not have, in general, a way for evaluating analytically the scaling of $\sigma$. 
However, since we aim to study the system in the vicinity of the critical point, we consider some reasonable assumptions for making the problem tractable. 
 For making progress, we notice that, since $\tau$ is an irrelevant parameter, its renormalized value goes to zero on the critical surface. Once we approach the critical point we can thus replace the correlated noise over $\tau$ with a white noise that is delta-correlated. \rev{It is worth noting that this approximation has to be reconsidered in the case of large $\tau$ values.} 
Moreover, as well as in the case of Active Ornstein-Uhlenbeck particles \cite{Fodor16}, we replace the average over trajectories with an average over the stationary distribution $P_s[\varphi]$, i.e.,  $\langle \mathcal{O}[\varphi] \rangle_{\eta} = \int \mathcal{D}[\varphi] P_s[\varphi] \mathcal{O}[\varphi]$. Again, because $\tau$ is an irrelevant parameter these averages can be done by replacing the stationary distribution with the Boltzmann distribution, i.e., $P_s[\varphi]\simeq e^{-\beta H_{LG}}$, with  the inverse temperature $\beta$ set to $1$ in the following. 
Once we perform these approximations, 
we obtain that $\sigma$ results from four contributions (in the following, the proportional symbol indicates that we are 
neglecting numerical factors that are inessential for our purpose)
\begin{align}
    \sigma   &\propto  \sigma_1 + \sigma_2 +   \sigma_3 + \sigma_4 \\ \nonumber
    \sigma_1 &\equiv \frac{\tau^2 u r^3}{D}     \langle \varphi^4 \rangle \nonumber \\
    \sigma_2 &\equiv \frac{\tau^2 u \mu^3}{D}   \langle (\nabla^2 \varphi)^3 \varphi  \rangle \nonumber \\ 
    \sigma_3 &\equiv \frac{\tau^2 r^2 \mu}{D}   \langle \varphi^3 \nabla^2 \varphi \rangle \nonumber \\ 
    \sigma_4 &\equiv \frac{\tau^2 u r \mu^2}{D} \langle (\nabla^2 \varphi)^2 \varphi^2 \rangle \; . \nonumber 
\end{align}
We first consider the four contributions above at the Gaussian level, i.e., by performing averages using the Gaussian measure given by $P_s = e^{-H_0}$. 
In the Gaussian model, we obtain that the only non-vanishing contribution in the limit $k \to 0$ is provided by $\sigma_1$. In particular, we get
\begin{align}
     \sigma_1 \propto \frac{\tau^2 u }{D} r^{1 + d/2} \; 
\end{align}
indicating that, at the mean-field level, i.e., above the upper critical dimension, the density of the entropy production rate tends to zero at the critical point, i.e., $\sigma \to 0$ for $r\to0$. More in general, by performing the naive Gaussian scaling analysis of $\sigma$, we obtain that it follows the natural scaling, i.e., $\sigma \to \sigma b^{-(d+z)}$, where the arrow indicates how $\sigma$ scales under Kadanoff transformations.
As discussed in Ref. \cite{PhysRevLett.124.240604} in the case of Active Model A, the fact that $\sigma$ goes to zero at the critical point does not guarantee that the dynamics is {\it effectively reversible} at criticality because what is really matter is the scaling of the singular part of the observable we are interested in. The suitable observable for doing that in the case of $\sigma$
is the density of the entropy production rate per spacetime correlation volume $\psi$ defined as $\psi \equiv \xi^{d+z} \sigma$ \cite{PhysRevLett.124.240604}. Since $\xi \sim r^{-\nu}$, and the mean-field values are $\nu_{MF}=1/2$ and $z_{MF}=2$, we obtain that $\psi = \xi^{d+2} \xi^{-\frac{1}{\nu_{MF}} \frac{d+2}{2} }=1$ meaning that, if we look for a singular part of $\psi$ that behaves as $\psi \sim r^{-\theta_\sigma}$, we get the mean-field value $\theta_\sigma=0$, which is in agreement also with Active Models A/B \cite{PhysRevLett.124.240604}. As discussed in detail in Ref. \cite{PhysRevLett.124.240604}, a non-negative value of $\theta_\sigma$, i.e., $\theta_\sigma \geq 0$, indicates already that, although the universality class of the model remains untouched, the dynamics remains irreversible even at the critical point.

We shall now provide arguments that support an exponent $\theta_{\sigma}$ of order $\epsilon$ around the upper critical dimension.
First of all, we notice that the only contribution to $\sigma$ that depends on a relevant operator is $\sigma_1$. In particular, the other composite operators in $\sigma_{2,3,4}$ have naive dimension $m+n>4$, with $m$ being the number of fields and $n$ being the number of gradients.
In order to understand the scaling of $\sigma_1$, we consider the following perturbative RG transformation (at the lowest order in $u$). Using perturbation theory, one has $\langle \mathcal{O}[\varphi] \rangle = \int \mathcal{D}[\varphi] \left( 1 - H_I \right) e^{-H_0} \mathcal{O}[\varphi]$ \cite{ma2018modern}. Since the observable we are going to consider is $\mathcal{O}=\varphi^4$, at the lowest order, one has $\sigma_1\!=\! \frac{\tau^2 u r^3}{D} \langle \varphi^4 \rangle_{H_0}$. 
Once we consider the scaling transformations (\ref{eq:scaling}), the standard one-loop RG equations bring us to the following scaling \cite{le1991quantum,tauber2014critical} (here we are using the fact that, under RG transformations, $r=r^\prime b^{-(z + \epsilon / 3)}$, $r \tau \!=\! r^\prime \tau^\prime \!=\! 1$ and $D\!=\!D^\prime\!=\!1$)
\begin{align}
    \sigma_1 =  b^{\epsilon/3}b^{-(z+d)} r^\prime u^\prime \langle (\varphi^\prime)^4 \rangle
\end{align}
with $\epsilon=4-d$. As a result, in this framework, the scaling of $\psi$ is $\psi \sim r^{-\theta_\sigma}$ with $\theta_\sigma=\frac{\nu \epsilon}{3}$ which is positive below the upper critical dimension.

\section*{Discussion and Conclusions}
In this work, motivated by the experimental and numerical evidence that active baths always develop correlations on finite time scales \cite{Maggi14,maggi2017memory,Maggi2021}, we have computed the entropy production rate in the case of a scalar field theory (with a non-conserved order parameter) driven out-of-equilibrium by a persistent noise. 
The computation predicts a non-vanishing entropy production rate only when non-linear interactions are taken into account (with a vanishing entropy production rate at equilibrium, i.e., that is recovered for $\tau\!=\!0$). 
We have performed a scaling analysis of the density of the entropy production rate $\sigma$ near the critical point in a static picture by replacing the averages over the persistent dynamics with averages over equilibrium configurations. 
The computation provides evidence in favor of an exponent $\theta_\sigma$ that is positive below the upper critical dimension and negative above it. However, even in this simplified framework, to have a better estimate of the critical exponent, the finest computations involving the scaling of composite operators are required \cite{amit2005field}. 
Moreover, it will be crucial to test those predictions against numerical data in three dimensions.
Since the non-linear terms responsible for a non-vanishing $\sigma$ in the model presented here are different from those in Active Model A, it is quite natural to obtain different critical scaling of $\sigma$. In Active Model A/B the leading non-equilibrium contribution is provided by a non-integrable gradient term, while the stochastic force at the coarse-grained level is considered equilibrium-like. Here we neglected non-equilibrium gradient terms and we focused our attention on the role of a non-equilibrium noise (as a step forward, it might be interesting to consider both contributions). 
It would be also interesting to investigate possible connections between the scaling of $\sigma$ and other thermodynamics anomalies observed at the microscale \cite{PhysRevLett.109.260603}.

As a future direction, the computation of $\sigma$ presented here might be extended to the case of Model B and in the presence of noise that is also correlated in space \cite{Maggi2021}. 
It might be also interesting to compare the phenomenological model presented here with coarse-graining descriptions of active systems obtained from microscopic models \cite{marconi2021hydrodynamics}. 

\section*{Acknowledgments}
I am deeply grateful to Mattia Scandolo (as well as to all the CoBBS group in Rome) 
for his comments and his critical reading of the manuscript. I also thank Claudio Maggi and Andrea Puglisi for illuminating discussions and for their critical reading of the early version of the manuscript.
This work has received funding from the European Union's Horizon 2020 research and innovation programme under the MSCA grant agreement No 801370
and by the Secretary of Universities 
and Research of the Government of Catalonia through Beatriu de Pin\'os program Grant No. BP 00088 (2018). 

\bibliography{mpbib}
\bibliographystyle{rsc}

\end{document}